\documentclass[12pt]{article}
\usepackage{wrapft}
\usepackage{bm}

\begin{document}

\title{$\mu$ Parameter from Dynamical Rearrangement of $U(1)$ and $\theta$ Parameter}

\author{
Yoshiharu \textsc{Kawamura}\footnote{E-mail: haru@azusa.shinshu-u.ac.jp} 
and Takashi \textsc{Miura}\footnote{E-mail: s09t302@shinshu-u.ac.jp}\\
{\it Department of Physics, Shinshu University, }\\
{\it Matsumoto 390-8621, Japan}
}

\date{
August 4, 2011}

\maketitle
\begin{abstract}
We study the generation of $\mu$ parameter from the dynamical rearrangement 
of local $U(1)$ symmetry in a five-dimensional model
and discuss phenomenological implications on the $\theta$ parameter,
under the assumption that supersymmetry is broken by the Scherk-Schwarz mechanism.
\end{abstract}

\section{Introduction}

The origin of soft supersymmetry (SUSY) breaking terms and $\mu$ term is
one of the biggest problem in the SUSY extension of standard model (SM).
It is usually expected that a high-energy physics is described by a quantum field theory (QFT) respecting SUSY, 
the SUSY is spontaneously broken in some hidden sector,
and soft SUSY breaking terms are induced in our visible sector by the mediation of some messengers.\cite{N}
Several mechanisms are proposed to generate the $\mu$ term with the breakdown of SUSY.\cite{K&N,mu1,mu2,mu3,mu4}

In our previous work, we proposed the mechanism that the soft SUSY breaking masses
and $\mu$ parameter can be induced from the dynamical rearrangement of local $U(1)$ symmetries
in a five-dimensional model, 
in quest of the possibility that SUSY is not completely realized in our starting high-energy QFT.\cite{K&M}\footnote{
Using the Hosotani mechanism, a similar dynamical generation of soft SUSY breaking masses was investigated 
in the supergravity (SUGRA) on the five-dimensional space-time 
including the orbifold $S^1/Z_2$ as the extra space.\cite{G&Q,GQ&R}
}
It is based on 
{\it a lesson from the brane world scenario such that symmetries are not necessarily realized uniformly
over the space-time including extra dimensions.}\cite{M&P,K1} 
As a weak point, our previous model has the low predictability for the magnitude of $\mu$ parameter.
This is because the $\mu$ parameter is given by the relevant extra $U(1)$ charge and 
the VEV of Wilson line phase which is determined by the one-loop effective potential
independent of the magnitude of soft SUSY breaking parameters.
Hence we hit on the idea that it would be improved if the one-loop effective potential
contains parameters relating soft SUSY breaking parameters and $\mu$ parameter at the same time.

In this paper, we study the generation of $\mu$ parameter from the dynamical rearrangement 
of local $U(1)$ symmetry in a five-dimensional model,
under the assumption that SUSY is broken by the Scherk-Schwarz mechanism.
The magnitude of $\mu$ parameter can be same as that of soft SUSY breaking masses of hidden scalar fields (${\phi_k}^l$)
because the one-loop effective potential includes both the Wilson line phase and the Scherk-Schwarz phases of ${\phi_k}^l$.
We discuss phenomenological implications on the $\theta$ parameter.
In the next section, we give a model and the one-loop effective potential
and derive the $\mu$ parameter.
In the last section, conclusions and a discussion are presented.

\section{Our model}

\subsection{Setup and outline}

We consider a SUSY extension of SM on the five-dimensional space-time $M^4 \times S^1/Z_2$.
The coordinates $x^{M}$ $(M =  0,1,2,3,5)$ are separated into 
the uncompactified four-dimensional ones $x^{\mu}$ $(\mu =  0,1,2,3)$ (or $x$)
and the compactified one $x^5$ (or $y$).
The $S^1/Z_2$ is obtained by dividing the circle $S^1$ (with the identification $y \sim y + 2\pi R$) 
by the $Z_2$ transformation $y \to -y$, so that the point $y$ is identified with $-y$.
Then the $S^1/Z_2$ is regarded as an interval with length $\pi R$, with $R$ being the $S^1$ radius.
Both end points $y=0$ and $\pi R$ are fixed points under the $Z_2$ transformation.
We regard the four-dimensional hypersurface on $y=0$ as our visible world.
 
We introduce the source of  SUSY breaking through the Scherk-Schwarz mechanism.\cite{S&S1,S&S2} 
It is applied to the SUSY SM on the five-dimensional space-time and 
the minimal SUSY extension of SM (MSSM) is derived on the brane.\cite{P&Q,BH&N1,BH&N2}
The magnitude of soft SUSY breaking masses ($m_{\tiny{\mbox{SUSY}}}$) is $O(\alpha_l/R)$
where $\alpha_l$ are twisted phases relating BCs and attached to the doublets of the $R$ symmetry $SU(2)_R$.
We refer to them as Scherk-Schwarz phases.
Tiny phases of $\alpha_l = O(R/\mbox{TeV}^{-1})$ are necessary (in the unit of TeV$^{-1}$ for the radius of extra dimension $R$)
in order to obtain $m_{\tiny{\mbox{SUSY}}}$ of $O(1)$TeV.

We bring extra $U(1)$ gauge symmetry which become a seed to the $\mu$ term
through the dynamical rearrangement, in place of the twisted phase relating the $\mu$ parameter.\footnote{
The $\mu$ parameter can be derived from the twisted phase relating BCs and attached to the doublets of $SU(2)_H$ for
Higgs hypermultiplets.\cite{P&Q,BH&N1,BH&N2}}
The dynamical rearrangement is a part of Hosotani mechanism.\cite{H1,H2,HHHK}
The physical symmetry and spectra are obtained after the determination of vacuum state via the mechanism.
An excellent feature is that the physics is mostly dictated by the particle contents of the theory
including the assignment of gauge quantum numbers.

When we utilize the Hosotani mechanism,
we encounter the problem that the fifth component of extra $U(1)$ gauge boson ($A_{5}$)
has usually $Z_2$ odd parities and it cannot play the role of Wilson line phase.
For the problem, there are two ways to make $Z_2$ parities of $A_5$ even.
One is that we impose the conjugate BCs on fields.\cite{HK&O}
This type of BCs are not suitable for the SM gauge bosons
and matter fields because zero modes of the SM gauge bosons are projected out
and zero modes of matter fields possess only real components.
But they can be useful for SM singlets.
The other is that we use a variant of the diagonal embedding proposed in Ref.~\cite{KTY}.
As shown later, we can impose appropriate BCs on both SM gauge bosons and matter fields, 
and the non-abelian structure such as $SU(2)$ can be build
by making a difference between the eigenstates of $U(1)$ gauge symmetry and those of BCs.
Hence we adopt the latter one for the MSSM fields
and the former one for SM singlet fields introduced in order to induce a non-vanishing Wilson line phase of extra $U(1)$ symmetry.

\subsection{MSSM particles}

First let us prepare bulk fields whose massless modes (in the absence of Schark-Schwarz phases)
at the tree level contain the MSSM particles.
The minimal sets are given with those BCs including the Scherk-Schwarz phases 
($\alpha_{\lambda}$, $\alpha_i$, $\alpha_h$) as follows.

~~\\
(i) The member of would-be MSSM gauge multiplets are $(A_M, \Sigma; \lambda^1, \lambda^2)$
whose BCs are given by
\begin{eqnarray}
\hspace{-1.2cm} &~& A_{M}(x,y+2\pi R)= A_{M}(x,y)~,~~ \Sigma(x,y+2\pi R)= \Sigma(x,y)~,~~ 
\nonumber \\
\hspace{-1.2cm} &~& 
\left(
\begin{array}{c}
\lambda^{1} \\ 
\lambda^{2}
\end{array}
\right)(x,y+2\pi R)= e^{2\pi i \alpha_{\lambda}\tau_2} 
\left(
\begin{array}{c}
\lambda^{1} \\ 
\lambda^{2}
\end{array}
\right)(x,y)~,
\label{Gauge-BC1}\\ 
\hspace{-1.2cm} &~& A_{\mu}(x,-y)= A_{\mu}(x,y)~,~~ 
\left(
\begin{array}{c}
A_{5} \\ 
\Sigma
\end{array}
\right)(x,-y)
= -\left(
\begin{array}{c}
A_{5} \\ 
\Sigma
\end{array}
\right)(x,y)~,~~ 
\nonumber \\
\hspace{-1.2cm} &~& 
\left(
\begin{array}{c}
\lambda^{1} \\ 
\lambda^{2}
\end{array}
\right)(x,-y)= \gamma_5 
\left(
\begin{array}{c}
-\lambda^{1} \\ 
\lambda^{2}
\end{array}
\right)(x,y)~,
\label{Gauge-BC2}\\
\hspace{-1.2cm} &~& A_{\mu}(x,2\pi R-y)= A_{\mu}(x,y)~,~~ 
\left(
\begin{array}{c}
A_{5} \\ 
\Sigma
\end{array}
\right)(x,2\pi R-y)
= -\left(
\begin{array}{c}
A_{5} \\ 
\Sigma
\end{array}
\right)(x,y)~,~~ 
\nonumber \\
\hspace{-1.2cm} &~& 
\left(
\begin{array}{c}
\lambda^{1} \\ 
\lambda^{2}
\end{array}
\right)(x,2\pi R-y)=  e^{2\pi i \alpha_{\lambda}\tau_2} \gamma_5
\left(
\begin{array}{c}
-\lambda^{1} \\ 
\lambda^{2}
\end{array}
\right)(x,y)~,
\label{Gauge-BC3}
\end{eqnarray}
where $A_M$ is the five-dimensional SM gauge bosons, $\Sigma$ is a real scalar field and $(\lambda^1, \lambda^2)$
are gauginos represented by symplectic-Majorana fermions.
The index indicating the SM gauge group or generators is suppressed.
The massless fields come from $A_{\mu}$ and $\lambda^1_{L}$.
The gauginos can acquire the $\alpha_{\lambda}$-dependent masses.

~~\\
(ii) The member of would-be MSSM matter multiplets are $(\psi^i; \phi^i, \phi^{ci\dagger})$
whose BCs are given by
\begin{eqnarray}
\hspace{-1.4cm} &~& \psi^i(x,y+2\pi R)= \psi^i(x,y)~,
\left(
\begin{array}{c}
\phi^i \\ 
\phi^{ci\dagger}
\end{array}
\right)\!\!(x,y+2\pi R)= e^{2\pi i \alpha_{i}\tau_2}
\left(
\begin{array}{c}
\phi^i \\ 
\phi^{ci\dagger}
\end{array}
\right)\!\!(x,y)~,~ 
\label{Matter-BC1}\\ 
\hspace{-1.4cm} &~& \psi^i(x, -y)= -\gamma_5 \psi^i(x,y)~,
\left(
\begin{array}{c}
\phi^i \\ 
\phi^{ci\dagger}
\end{array}
\right)\!\!(x, -y)= 
\left(
\begin{array}{c}
\phi^i \\ 
-\phi^{ci\dagger}
\end{array}
\right)\!\!(x,y)~,
\label{Matter-BC2}\\
\hspace{-1.4cm} &~& \psi^i(x, 2\pi R-y)= -\gamma_5 \psi^i(x,y)~,
\nonumber \\
\hspace{-1.4cm} &~& 
\left(
\begin{array}{c}
\phi^i \\ 
\phi^{ci\dagger}
\end{array}
\right)\!\!(x, 2\pi R-y)= e^{2\pi i \alpha_{i}\tau_2}
\left(
\begin{array}{c}
\phi^i \\ 
-\phi^{ci\dagger}
\end{array}
\right)\!\!(x,y)~,
\label{Matter-BC3}
\end{eqnarray}
where $\psi^i$ are fermions represented by four-component spinors and $(\phi^i, \phi^{ci\dagger})$ are complex scalar fields.
The index $i$ represents particle species.
The massless fields come from $\psi^i_L$ and $\phi^i$
which are the chiral fermions (quarks and leptons) and the corresponding scalar bosons (squarks and sleptons), respectively.
The scalar bosons can acquire the $\alpha_{i}$-dependent masses.

~~\\
(iii) The member of would-be MSSM Higgs multiplets are $(\tilde{h}; h,h^{c\dagger})$ and $(\tilde{\bar{h}}; \bar{h},\bar{h}^{c\dagger})$.
The MSSM Higgsinos come from the fermions $\tilde{h}$ and $\tilde{\bar{h}}$.
The MSSM Higgs bosons stem from the complex scalar fields $(h,h^{c})$ and $(\bar{h},\bar{h}^{c})$.
We use the notation such that $\tilde{h}_1 = \tilde{h}$, $\tilde{h}_2 = \tilde{\bar{h}}^{\dagger}$ for Higgsinos
and ${h_1}^1 = h$, ${h_1}^2 = h^{c\dagger}$, ${h_2}^1 = \bar{h}^{\dagger}$, ${h_2}^2 = \bar{h}^{c}$ for Higgs bosons.
We impose the following BCs on Higgsinos,
\begin{eqnarray}
\hspace{-1.4cm} &~& 
\left(
\begin{array}{c}
\tilde{h}_1 \\ 
\tilde{h}_2
\end{array}
\right)\!\!(x,y+2\pi R)= 
\left(
\begin{array}{c}
\tilde{h}_1 \\ 
\tilde{h}_2
\end{array}
\right)\!\!(x,y)~,~ 
\label{Higgsino-BC1}\\ 
\hspace{-1.4cm} &~& 
\left(
\begin{array}{c}
\tilde{h}_1 \\ 
\tilde{h}_2
\end{array}
\right)\!\!(x, -y)= \gamma_5
\left(
\begin{array}{c}
\tilde{h}_1 \\ 
-\tilde{h}_2
\end{array}
\right)\!\!(x,y)~,
\label{Higgsino-BC2}\\
\hspace{-1.4cm} &~& 
\left(
\begin{array}{c}
\tilde{h}_1 \\ 
\tilde{h}_2
\end{array}
\right)\!\!(x, 2\pi R-y)= \gamma_5
\left(
\begin{array}{c}
\tilde{h}_1 \\ 
-\tilde{h}_2
\end{array}
\right)\!\!(x,y)
\label{Higgsino-BC3}
\end{eqnarray}
and on Higgs bosons,
\begin{eqnarray}
\hspace{-1.4cm} &~& 
\left(
\begin{array}{c}
{h_1}^1 \\ 
{h_1}^2
\end{array}
\right)\!\!(x,y+2\pi R)= 
e^{2\pi i \alpha_h \tau_2}
\left(
\begin{array}{c}
{h_1}^1 \\ 
{h_1}^2
\end{array}
\right)\!\!(x,y)~,~~
\nonumber \\
\hspace{-1.4cm} &~& 
\left(
\begin{array}{c}
{h_2}^1 \\ 
{h_2}^2
\end{array}
\right)\!\!(x,y+2\pi R)= 
e^{2\pi i \alpha_h \tau_2}
\left(
\begin{array}{c}
{h_2}^1 \\ 
{h_2}^2
\end{array}
\right)\!\!(x,y)~,
\label{Higgs-BC1}\\ 
\hspace{-1.4cm} &~& 
\left(
\begin{array}{c}
{h_1}^1 \\ 
{h_1}^2
\end{array}
\right)\!\!(x, -y)=
\left(
\begin{array}{c}
{h_1}^1 \\ 
-{h_1}^2
\end{array}
\right)\!\!(x,y)~,
\nonumber \\
\hspace{-1.4cm} &~& 
\left(
\begin{array}{c}
{h_2}^1 \\ 
{h_2}^2
\end{array}
\right)\!\!(x, -y)=
\left(
\begin{array}{c}
-{h_2}^1 \\ 
{h_2}^2
\end{array}
\right)\!\!(x,y)~,
\label{Higgs-BC2}\\
\hspace{-1.4cm} &~& 
\left(
\begin{array}{c}
{h_1}^1 \\ 
{h_1}^2
\end{array}
\right)\!\!(x,2\pi R-y)= 
e^{2\pi i \alpha_h \tau_2}
\left(
\begin{array}{c}
{h_1}^1 \\ 
-{h_1}^2
\end{array}
\right)\!\!(x,y)~,~~
\nonumber \\
\hspace{-1.4cm} &~& 
\left(
\begin{array}{c}
{h_2}^1 \\ 
{h_2}^2
\end{array}
\right)\!\!(x,2\pi R-y)= 
e^{2\pi i \alpha_h \tau_2}
\left(
\begin{array}{c}
-{h_2}^1 \\ 
{h_2}^2
\end{array}
\right)\!\!(x,y) ~.
\label{Higgs-BC3}
\end{eqnarray}
~~\\

In the bulk, there are no Yukawa couplings among matter multiplets and Higgs multiplets because of SUSY.
We impose the $R$ symmetry on the model.
Then the theory on our brane is described by the SUSY Lagrangian of the MSSM, using the above particle contents.
The $\mu$ term is forbidden at the tree level by the $R$ symmetry.
Upon compactification, the following mass terms are derived,\cite{P&Q,BH&N1,BH&N2}
\begin{eqnarray}
\mathcal{L}_{\tiny{\mbox{soft}}} = -\left(\frac{1}{2} \sum_{a} M_{\lambda}^a {\lambda}^a \lambda^a + \mbox{h.c.}\right)
- \sum_i m_i^2 |\phi^i|^2 - m_{h_u}^2 |h_u|^2 - m_{h_d}^2 |h_d|^2~,
\label{Lsoft}
\end{eqnarray}
where $\lambda^a$ are the MSSM gauginos, $\phi^i$ are the MSSM sfermions
and $(h_u, h_d)$ are the MSSM Higgs bosons,
and $M_{\lambda^a}$, $m_i$, $m_{h_u}$ and $m_{h_d}$ are the soft SUSY breaking masses given by
\begin{eqnarray}
M_{\lambda^a} = \frac{\alpha_{\lambda}}{R}~,~~ m_i^2 = \left(\frac{\alpha_i}{R}\right)^2~,~~
m_{h_u}^2 = m_{h_d}^2 = \left(\frac{\alpha_h}{R}\right)^2~.
\label{softM}
\end{eqnarray}
Note that $\mu$ and $B$ terms do not appear
in the absence of the twisted phase attached to the doublets of $SU(2)_H$ for Higgs hypermultiplets.

\subsection{Extra $U(1)$ symmetry}

We introduce extra $U(1)$ gauge symmetry to generate $\mu$ and $B$ terms through 
the dynamical rearrangement.
Let the abelian gauge boson ${A'}_M^{(-)}$ of $U(1)'^{(-)}$
satisfy the BCs:
\begin{eqnarray}
\hspace{-1cm} &~& {A'}_M^{(-)}(x,y+2\pi R)={A'}_M^{(-)}(x,y)~,~~
\label{Apm-BC1}\\
\hspace{-1cm} &~& {A'}_{\mu}^{(-)}(x,-y)=- {A'}_{\mu}^{(-)}(x,y)~,~~ {A'}_{5}^{(-)}(x,-y)= {A'}_{5}^{(-)}(x,y)~,
\label{Apm-BC2}\\
\hspace{-1cm} &~& {A'}_{\mu}^{(-)}(x, 2\pi R-y)=- {A'}_{\mu}^{(-)}(x,y)~,~~ {A'}_{5}^{(-)}(x,2\pi R-y)= {A'}_{5}^{(-)}(x,y)~.
\label{Apm-BC3}
\end{eqnarray}
{}From (\ref{Apm-BC1}) -- (\ref{Apm-BC3}),
we find that the $U(1)'^{(-)}$ is broken down on our brane
because the massless mode do not appear in ${A'}_{\mu}^{(-)}$ but ${A'}_{5}^{(-)}$.
The massless mode of ${A'}_{5}^{(-)}$ is a dynamical field which will play a central role as the Wilson line phase
in the dynamical rearrangement.\footnote{
The superpartners $\lambda'^{(-)}_1$ and $\lambda'^{(-)}_2$ of $U(1)'^{(-)}$ gauge boson are also introduced.
They cannot couple to matter multiplets but Higgs multiplets.
We do not discuss their phenomenological implications because the mass spectra depend on the BCs.
}

Let us assume that the interactions of ${A'}_M^{(-)}$ and Higgsinos are described by the Lagrangian density
\begin{eqnarray}
\mathcal{L}_{\tilde{h}} = i (\overline{\tilde{h}}_1, \overline{\tilde{h}}_2) \Gamma^M
\left(
\begin{array}{cc}
\partial_M  & ig'_{\tiny{\mbox{5D}}} q'_{h}{A'}_M^{(-)} \\
ig'_{\tiny{\mbox{5D}}} q'_{h}{A'}_M^{(-)} & \partial_M 
\end{array}
\right)
\left(
\begin{array}{c}
\tilde{h}_1 \\
\tilde{h}_2
\end{array}
\right)~,
\label{tildeh-L}
\end{eqnarray}
where $g'_{\tiny{\mbox{5D}}}$ is the gauge coupling of $U(1)'^{(-)}$ with mass dimension $[g'_{\tiny{\mbox{5D}}}] = -1/2$.
Here and hereafter we omit the SM gauge bosons irrelevant of our discussion
to avoid a complication.
Note that the $\tilde{h}_k$ ($k=1,2$) are the eigenstates of BCs,
but they are not the eigenstates of $U(1)'^{(-)}$ symmetry.
If ${A'}_5^{(-)}$ acquires a non-vanishing vacuum expectation value (VEV), 
the following Higgsino mass term is induced
\begin{eqnarray}
\mathcal{L}_{\tilde{h}}^{\rm mass} = - \left(\mu {\tilde{h}}_u \tilde{h}_d+ \mbox{h.c.}\right)~,
\label{Lmu}
\end{eqnarray}
where $\mu = g'_{\tiny{\mbox{5D}}}q'_{h}\langle {A'}_5^{(-)} \rangle$
and $(\tilde{h}_u, \tilde{h}_d)$ are the MSSM Higgsinos.

The eigenstates of $U(1)'^{(-)}$ symmetry can be constructed as a linear combination such that
\begin{eqnarray}
\tilde{H}_1 = \frac{\tilde{h}_1 + \tilde{h}_2}{\sqrt{2}}~,~~ \tilde{H}_2 = \frac{\tilde{h}_1 - \tilde{h}_2}{\sqrt{2}}~.
\label{tildeH}
\end{eqnarray}
In fact, $\mathcal{L}_{\tilde{h}}$ is rewritten by
\begin{eqnarray}
\mathcal{L}_{\tilde{h}} =
i (\overline{\tilde{H}}_1, \overline{\tilde{H}}_2) \Gamma^M
\left(
\begin{array}{cc}
\partial_M + ig'_{\tiny{\mbox{5D}}} q'_{h}{A'}_M^{(-)} & 0\\
0 & \partial_M - ig'_{\tiny{\mbox{5D}}} q'_{h}{A'}_M^{(-)} 
\end{array}
\right)
\left(
\begin{array}{c}
\tilde{H}_1 \\
\tilde{H}_2
\end{array}
\right)~,
\label{tildeH-L}
\end{eqnarray}
where $q'_h$ is the charge of $U(1)'^{(-)}$ with the mass dimension $[q'_h] = 0$.
We need a pair of fields whose $U(1)'^{(-)}$ charge has an opposite value, i. e. $\tilde{H}_1$ and $\tilde{H}_2$.
They satisfy the BCs:
\begin{eqnarray}
&~& \tilde{H}_k (x,y+2\pi R)= \tilde{H}_k (x,y)~,~~
\tilde{H}_1 (x,-y)= \gamma_5 \tilde{H}_2 (x,y)~,~~
\nonumber \\
&~& \tilde{H}_1 (x,2\pi R -y)= \gamma_5 \tilde{H}_2 (x,y)
\label{tildeHBCs}
\end{eqnarray}
and are expanded as
\begin{eqnarray}
\hspace{-0.7cm} &~& \tilde{H}_1(x,y)= \sum_{n = -\infty}^{\infty} \tilde{H}_n(x) \exp\left(i\frac{n}{R} y\right)~,
\label{Psi-1-exp}\\ 
\hspace{-0.7cm} &~& \tilde{H}_2(x,y)= \sum_{n = -\infty}^{\infty} \gamma_5 \tilde{H}_n(x) \exp\left(-i\frac{n}{R} y\right)~,
\label{Psi-2-exp}
\end{eqnarray}
respectively.
Here and hereafter a common normalization factor is omitted.
Upon compactification, the following mass terms appear after integrating over $y$,
\begin{eqnarray}
{\sum_{n = -\infty}^{\infty}} \frac{n + q'_h \gamma}{R} 
\left(\overline{\tilde{H}}_{n{\rm L}}(x) \tilde{H}_{n{\rm R}}(x) + \overline{\tilde{H}}_{n{\rm R}}(x) \tilde{H}_{n{\rm L}}(x)\right)~,
\label{psi-mass}
\end{eqnarray}
where $\gamma \equiv g'_{\tiny{\mbox{5D}}} \langle {A'}_5^{(-)} \rangle R$ with the mass dimension $[\gamma] = 0$
and $i$ is absorbed into the phases of fields.
{}From (\ref{psi-mass}), the following one-loop effective potential is obtained
\begin{eqnarray}
V_{\rm eff}^{\tilde{h}}[\gamma] = 8C\sum_{n=1}^{\infty}{1 \over n^5}\cos{\left[2\pi n\left(q'_{h}\gamma\right)\right]}~,
\label{Psi-Veff}
\end{eqnarray}
where $C\equiv 3/(128\pi^6 R^4)$.
Here and hereafter $\gamma$-independent terms are omitted.

As the same way, let us assume that the interactions of ${A'}_M^{(-)}$ and Higgs bosons are described 
by the Lagrangian density
\begin{eqnarray}
&~& \mathcal{L}_{h} = 
\left|\left(
\begin{array}{cc}
\partial_M & ig'_{\tiny{\mbox{5D}}} q'_{h}{A'}_M^{(-)} \\
 ig'_{\tiny{\mbox{5D}}} q'_{h}{A'}_M^{(-)} & \partial_M  
\end{array}
\right)
\left(
\begin{array}{c}
{h_1}^1 \\
{h_2}^1 
\end{array}
\right) \right|^2
\nonumber \\
&~& ~~~~~~~~ +
\left|\left(
\begin{array}{cc}
\partial_M & ig'_{\tiny{\mbox{5D}}} q'_{h}{A'}_M^{(-)} \\
ig'_{\tiny{\mbox{5D}}} q'_{h}{A'}_M^{(-)}  & \partial_M 
\end{array}
\right)
\left(
\begin{array}{c}
{h_1}^2 \\
{h_2}^2 
\end{array}
\right) \right|^2~.
\label{h-L}
\end{eqnarray}
Note that the ${h_k}^l$  ($k=1,2$, $l=1,2$) are not the eigenstates of $U(1)'^{(-)}$ symmetry.
If ${A'}_5^{(-)}$ acquires a non-vanishing VEV, 
the following Higgs mass term is induced
\begin{eqnarray}
\mathcal{L}_{h}^{\rm mass} = -\left(\mu^2 + \left(\frac{\alpha_h}{R}\right)^2\right) \left(|h_u|^2 + |h_d|^2\right) - B\mu\left( h_u h_d + \mbox{h.c.}\right)~,
\label{LmuB}
\end{eqnarray}
where $B\mu = 2\alpha_h g'_{\tiny{\mbox{5D}}} q'_{h}\langle {A'}_5^{(-)} \rangle/R =2\alpha_h q'_{h}\gamma/R^2$
and $i$ is absorbed into the phases of fields.

The eigenstates of $U(1)'^{(-)}$ symmetry can be constructed as a linear combination such that
\begin{eqnarray}
&~& H_1 = \frac{{h_1}^1 + {h_2}^1}{\sqrt{2}}~,~~ {H_2} = \frac{{h_1}^1 - {h_2}^1}{\sqrt{2}}~,~
\nonumber \\
&~& {H_3} = \frac{{h_1}^2 + {h_2}^2}{\sqrt{2}}~,~~ {H_4} = \frac{-{h_1}^2 + {h_2}^2}{\sqrt{2}}~.
\label{H}
\end{eqnarray}
In fact, $\mathcal{L}_{h}$ is rewritten by
\begin{eqnarray}
&~& \mathcal{L}_{h} = \left|\left(\partial_M + ig'_{\tiny{\mbox{5D}}} q'_{h}{A'}_M^{(-)}\right)H_1\right|^2
+ \left|\left(\partial_M - ig'_{\tiny{\mbox{5D}}} q'_{h}{A'}_M^{(-)}\right)H_2\right|^2
\nonumber \\
&~& ~~~~ + \left|\left(\partial_M + ig'_{\tiny{\mbox{5D}}} q'_{h}{A'}_M^{(-)}\right) H_3\right|^2
+ \left|\left(\partial_M - ig'_{\tiny{\mbox{5D}}} q'_{h}{A'}_M^{(-)}\right)H_4\right|^2~.
\label{H-L}
\end{eqnarray}
The $H_a$ ($a=1,2,3,4$) satisfy the BCs:
\begin{eqnarray}
\hspace{-1.4cm} &~& 
\left(
\begin{array}{c}
H_1 \\ 
H_3
\end{array}
\right)\!\!(x,y+2\pi R)= 
e^{2\pi i \alpha_h \tau_2}
\left(
\begin{array}{c}
H_1 \\ 
H_3
\end{array}
\right)\!\!(x,y)~,~~
\nonumber \\
\hspace{-1.4cm} &~& 
\left(
\begin{array}{c}
H_2 \\ 
H_4
\end{array}
\right)\!\!(x,y+2\pi R)= 
e^{-2\pi i \alpha_h \tau_2}
\left(
\begin{array}{c}
H_2 \\ 
H_4
\end{array}
\right)\!\!(x,y) ~,
\label{Higgs-BC1-H}\\ 
\hspace{-1.4cm} &~& 
{H_1}(x, -y) = {H_2}(x, y)~,~~ {H_3}(x, -y) = {H_4}(x, y)~,
\label{Higgs-BC2-H}\\
\hspace{-1.4cm} &~& 
\left(
\begin{array}{c}
{H_1} \\ 
{H_3}
\end{array}
\right)\!\!(x,2\pi R-y)= 
e^{2\pi i \alpha_h \tau_2}
\left(
\begin{array}{c}
{H_2} \\ 
{H_4}
\end{array}
\right)\!\!(x,y)
\label{Higgs-BC3-H}
\end{eqnarray}
and are expanded as
\begin{eqnarray}
\hspace{-0.6cm}&~& H_1(x, y) = \sum_{n = -\infty}^{\infty} \left[h_n^{u}(x) \cos \frac{n-\alpha_h}{R} y 
 + h_n^{d\dagger}(x) \sin \frac{n-\alpha_h}{R} y\right]~,
\label{H1-exp}\\
\hspace{-0.6cm}&~& H_2(x, y) = \sum_{n = -\infty}^{\infty} \left[h_n^{u}(x) \cos \frac{n-\alpha_h}{R} y 
 - h_n^{d\dagger}(x) \sin \frac{n-\alpha_h}{R} y\right]~,
\label{H2-exp}\\
\hspace{-0.6cm}&~& H_3(x, y) = \sum_{n = -\infty}^{\infty} \left[h_n^{u}(x) \sin \frac{n-\alpha_h}{R} y 
 - h_n^{d\dagger}(x) \cos \frac{n-\alpha_h}{R} y\right]~,
\label{H3-exp}\\
\hspace{-0.6cm}&~& H_4(x, y) = \sum_{n = -\infty}^{\infty} \left[-h_n^{u}(x) \sin \frac{n-\alpha_h}{R} y 
 - h_n^{d\dagger}(x) \cos \frac{n-\alpha_h}{R} y\right]~,
\label{H4-exp}
\end{eqnarray}
respectively.
Upon compactification, the following mass terms appear after integrating over $y$,
\begin{eqnarray}
&~& \sum_{n = -\infty}^{\infty} \left[ \left(\frac{n - \alpha_h}{R}\right)^2 +  \left(\frac{q'_h \gamma}{R}\right)^2 \right]
\left(|h_n^{u}|^2 + |h_n^{d}|^2\right)
\nonumber \\
&~&  ~~~~~~ + \sum_{n = -\infty}^{\infty} \left[2i \frac{n - \alpha_h}{R} \frac{q'_h \gamma}{R} h_n^{u} h_n^{d} + \mbox{h.c.}\right]~.
\label{hud-mass}
\end{eqnarray}
{}From (\ref{hud-mass}), we find that the eigenvalues of mass squared matrix for $h_n^{u}$ and $h_n^{d\dagger}$ 
are $\displaystyle{\frac{n - \alpha_h \pm q'_h \gamma}{R}}$ and the following one-loop effective potential is obtained
\begin{eqnarray}
V_{\rm eff}^{h}[\gamma] = -4C\sum_{n=1}^{\infty}{1 \over n^5} \left[\cos{\left[2\pi n\left(q'_{h}\gamma - \alpha_h \right)\right]}
 + \cos{\left[2\pi n\left(q'_{h}\gamma + \alpha_h \right)\right]}\right]~.
\label{Psi-Veff}
\end{eqnarray}

The sum of one-loop effective potentials $V_{\rm eff}^{\tilde{h}}[\gamma]$ and $V_{\rm eff}^{h}[\gamma] $ is rewritten by
\begin{eqnarray}
V_{\rm eff}^{H}[\gamma] = V_{\rm eff}^{\tilde{h}}[\gamma] + V_{\rm eff}^{h}[\gamma]
= 8C \sum_{n=1}^{\infty}{1 \over n^5} \left[1 - \cos(2\pi n \alpha_h)\right] \cos(2\pi n q'_{h}\gamma)~.
\label{Veff-total}
\end{eqnarray}
We find that the $V_{\rm eff}^{H}[\gamma]$ minimizes at $q'_h \gamma = 1/2$ irrespective to the value of $\alpha_h$.
This means that the magnitude of $\mu$ is $1/(2R)$ comparable to those of Kaluza-Klein modes
and the MSSM with soft SUSY breaking parameters of $O(1)$TeV and a small extra dimension much less than 
$O(10^{-18})$m cannot be derived unless other contributions are added.

\subsection{Fixation of Wilson line phase}

We introduce SM gauge singlets in order to fix $\gamma$ to be $O(R/\mbox{TeV}^{-1})$.   
Let us incorporate two hypermultiplets $(\psi_1; {\phi_1}^1, {\phi_1}^2)$ and $(\psi_2; {\phi_2}^1, {\phi_2}^2)$
whose Lagrangian density is given by
\begin{eqnarray}
\mathcal{L} = \sum_{k=1, 2} i \overline{\psi}_k \Gamma^M D_M \psi_k 
 + \sum_{k=1,2}\sum_{l=1,2} |D_M {\phi_k}^l|^2~,
\label{PsiPhi-L}
\end{eqnarray}
where $D_M \equiv \partial_M + i g'_{\tiny{\mbox{5D}}} q'_{\phi} {A'}_M^{(-)}$.
Those fields satisfy the conjugate BCs such that
\begin{eqnarray}
\hspace{-1.4cm} &~& 
\left(
\begin{array}{c}
\psi_1 \\ 
\psi_2
\end{array}
\right)\!\!(x,y+2\pi R)= 
- \left(
\begin{array}{c}
\psi_1 \\ 
\psi_2
\end{array}
\right)\!\!(x,y)~,~ 
\label{Psi-BC1}\\ 
\hspace{-1.4cm} &~& 
\left(
\begin{array}{c}
\psi_1 \\ 
\psi_2
\end{array}
\right)\!\!(x, -y)= \gamma_5
\left(
\begin{array}{c}
\psi_1^{\dagger} \\ 
-\psi_2^{\dagger}
\end{array}
\right)\!\!(x,y)~,
\label{Psi-BC2}\\
\hspace{-1.4cm} &~& 
\left(
\begin{array}{c}
\psi_1 \\ 
\psi_2
\end{array}
\right)\!\!(x, 2\pi R-y)= -\gamma_5
\left(
\begin{array}{c}
\psi_1^{\dagger} \\ 
-\psi_2^{\dagger}
\end{array}
\right)\!\!(x,y)
\label{Psi-BC3}
\end{eqnarray}
and
\begin{eqnarray}
\hspace{-1.4cm} &~& 
{\phi_k}^l(x,y+2\pi R)= e^{2\pi i \alpha_{\phi}} {\phi_k}^l(x,y)~,~~
{\phi_k}^l(x,-y)= {\phi_k}^{l\dagger}(x,y)~,~~
\nonumber \\
\hspace{-1.4cm} &~&
{\phi_k}^l(x,2\pi R-y)= e^{2\pi i \alpha_{\phi}} {\phi_k}^{l\dagger}(x,y)~,
\label{Phi-BC}
\end{eqnarray}
where we take a universal Scherk-Schwarz phase $\alpha_{\phi}$ for ${\phi_k}^l$.
The $\psi_1$ and $\psi_2$ are expanded as
\begin{eqnarray}
\hspace{-1.4cm} &~&
\psi_1(x, y) = \gamma_5 \sum_{n = 1}^{\infty} \psi_{1n}(x) \cos \left(\frac{n + \frac{1}{2}}{R} y\right)~,
\label{Psi-1-exp}\\
\hspace{-1.4cm} &~&
\psi_2(x, y) = \gamma_5 \sum_{n = 1}^{\infty} \psi_{2n}(x) \sin \left(\frac{n + \frac{1}{2}}{R} y\right)~,
\label{Psi-2-exp}
\end{eqnarray}
respectively.
Here $\psi_{1n}(x)$ and $\psi_{2n}(x)$ are real fields.
The ${\phi_k}^l$ are expanded as
\begin{eqnarray}
{\phi_k}^l(x, y) = \sum_{n = -\infty}^{\infty} \phi_{kn}^l(x) \exp \left(i\frac{n + \alpha_{\phi}}{R} y\right)~,
\label{Phi-exp}
\end{eqnarray}
where $\phi_{kn}^l(x)$ are real fields.
Using the above expansions (\ref{Psi-1-exp}) -- (\ref{Phi-exp}),
the following one-loop effective potential is derived
\begin{eqnarray}
V_{\rm eff}^{\phi}[\gamma]
 = 8C \sum_{n=1}^{\infty}{1 \over n^5}\left[\cos{\left[2\pi n \left({1\over 2} +q'_{\phi}\gamma\right)\right]} 
- \cos{\left[2\pi n \left(\alpha_{\phi} + q'_{\phi}\gamma\right)\right]}\right]~.
\label{Hid-Veff}
\end{eqnarray}

We consider the sum of one-loop effective potential for Higgs multiplet $V_{\rm eff}^{H}[\gamma]$ and 
the SM singlets one $V_{\rm eff}^{\phi}[\gamma]$. 
When $|\alpha_{\phi}| \ll 1$, the minimum of $V_{\rm eff}^{H}[\gamma] + V_{\rm eff}^{\phi}[\gamma]$ is given 
at $\displaystyle{\gamma \simeq -\frac{\alpha_{\phi}}{2q'_{\phi}}}$.
In the case that $\alpha_{\phi}/q'_{\phi} = O(\alpha_h/q'_h)$, 
the magnitude of $\mu = q'_{h} \gamma/R$ can be same order of soft SUSY breaking Higgs mass $|\alpha_{h}/R|$
and it offers to a solution of $\mu$ problem.
We find that the breakdown of electroweak symmetry can occur radiatively using the same analysis in \cite{BH&N1}.
The light SM gauge singlet fields with mass $|\alpha_{\phi}/R|$ appear upon compactification,
but it would be harmless from the viewpoint of naturalness 
unless they couple to Higgs multiplets through renormalizable interactions.\cite{NS&W,L,E,B&P}

Our model does not suffer from the SUSY CP problem because all soft SUSY breaking masses and $\mu$ parameter
are real-valued.
The flavor changing neutral current (FCNC) processes can be suppressed enough
if the Schark-Schwarz phase of the relevant sfermions has a common value.

Our scenario can be applied to the SUSY $SU(5)$ orbifold grand unified theory (GUT).\cite{K2,H&N}
By the introduction of the following matrices $(P_0, P_1)$ relating $Z_2$ reflections under $y=0$ and $y=\pi R$
to the BCs of two sets of Higgs multiplets,
\begin{eqnarray}
P_0 = \mbox{diag}(1,1,1,1,1)~,~~ P_1 = \mbox{diag}(-1,-1,-1,1,1)~,
\label{P0P1}
\end{eqnarray}
the triplet-doublet splittings are realized elegantly.

\subsection{Implication on $\theta$ parameter}

The strong CP problem is a naturalness problem that asks why the CP-violating phase in QCD is extremely small.\cite{strongCP1,strongCP2}
The non-observation of the neutron electric dipole moment suggests $|\bar{\theta}| \le O(10^{-10})$.\footnote{
After the breakdown of $SU(2)_L \times U(1)_Y$ and the re-definition of quark fields' phase,
the parameter $\theta$ becomes the effective one $\bar{\theta} \equiv \theta + {\mbox{argdet}}(M_u M_d)$
where $M_{u,d}$ are mass matrices of the up and down-type quarks.}
The parameter $\bar{\theta}$ is a physical one unless there is an exact global symmetry that can make $\bar{\theta}$ to be zero,
in which case the value of $\bar{\theta}$ is determined dynamically
by introducing a corresponding physical degrees of freedom.

Three possible solutions have been proposed to solve the strong CP problem.
First one is that one of quarks is massless and then $\bar{\theta}$ is made to be zero by the chiral transformation. 
This possibility seems to be ruled out by experiments.
Second one is the so-called Peccei-Quinn mechanism~\cite{PQ1,PQ2} 
involving a light pseudo Nambu-Goldstone boson called axion.\cite{axion1,axion2,axion3,axion4,axion5}
In the model, the Peccei-Quinn symmetry $U(1)_{\rm PQ}$ couples to the QCD anomaly 
and $\bar{\theta}$ is made to be zero dynamically by the potential generated by the QCD instanton effects.
Third one is that the CP transformation is an exact symmetry in an underlying high-energy theory, 
and it is broken very weakly in the low-energy theory.\cite{exactCP1,exactCP2}

The Peccei-Quinn mechanism is most popular, but there are two theoretical problems.
One is how to suppress contributions from other explicit $U(1)_{\rm PQ}$ breaking terms
such as higher-dimensional operators induced by the possible quantum gravity effects.
The other is how to get the axion decay constant $f_a$ naturally within the narrow window $f_a = 10^{10 \sim 12}$GeV, 
where the constraint on $f_a$ originates from astrophysical and cosmological bounds.

Let us discuss the implications on the $\theta$ parameter from the dynamical rearrangement.\footnote{
The dynamical rearrangement of the $\theta$ parameter has been studied in the presence of a mixed Chern-Simons term
and it has been pointed out that the parameter is regarded as a BCs for the orbifolding.\cite{HK&O}}
In our model, there are at least two sources to the strong CP violation.
The first one is the $\theta$ term on the brane,
\begin{eqnarray}
\mathcal{L}_{\rm brane}^{\theta} = \frac{\theta}{32\pi^2} \varepsilon^{\mu\nu\rho\sigma}\mbox{tr} \left(F_{\mu\nu}^{(0)} F_{\rho\sigma}^{(0)}\right)  ,
\label{Lbrane}
\end{eqnarray}
where $\theta$ is the QCD vacuum angle on the brane with the mass dimension $[\theta] = 0$.
The second one is the bulk term called the mixed Chern-Simons term,
\begin{eqnarray}
\mathcal{L}_{\rm bulk}^{\rm MCS} = {\kappa_{\tiny{\mbox{5D}}} \over 5!} \varepsilon^{MNLOP} A'^{(-)}_M \mbox{tr} \left(F_{NL}F_{OP}\right) ,
\label{LBulk}
\end{eqnarray}
where $\kappa_{\tiny{\mbox{5D}}}$ is the coupling constant of mass dimension $[\kappa_{\tiny{\mbox{5D}}}]=-1/2$. 
The mixed Chern-Simons term is invariant under the gauge transformation whose gauge function vanish at the boundaries of space-time.
Note that we have put the $\theta$ term only for the zero mode $F_{\mu\nu}^{(0)}$ 
though there can be other terms such as
$\displaystyle{\theta_n \varepsilon^{\mu\nu\rho\sigma} \mbox{tr} \left(F_{\mu\nu}^{(n)} F_{\rho\sigma}^{(n)}\right)}$. 

After the dimensional reduction and dynamical rearrangement,
we obtain the following QCD vacuum angle,
\begin{eqnarray}
\theta_{\rm eff} = \theta + \frac{4 \pi^2}{3} \frac{\kappa_{\tiny{\mbox{5D}}} \gamma}{g'_{\tiny{\mbox{5D}}} R} 
= \theta + \frac{4 \pi^2}{3} \frac{\kappa_{\tiny{\mbox{4D}}} \gamma}{g'_{\tiny{\mbox{4D}}}}~,
\label{thetaeff}
\end{eqnarray}
where $\kappa_{\tiny{\mbox{4D}}}$ and $g'_{\tiny{\mbox{4D}}}$ are couplings with mass dimesion zero
relating to $\kappa_{\tiny{\mbox{4D}}} = \kappa_{\tiny{\mbox{5D}}}/\sqrt{R}$
and $g'_{\tiny{\mbox{4D}}} = g'_{\tiny{\mbox{5D}}} \sqrt{R}$, respectively.

As pointed out in \cite{HK&O}, the ${A'}_5^{(-)}$ cannot play the role of axion
in the presence of one-loop effective potential $V_{\rm eff}[\gamma]$
because $V_{\rm eff}$ usualy gives the dominant contribution to determine the VEV of ${A'}_5^{(-)}$ 
against the QCD instanton effect.\footnote{
It was pointed out that the ${A'}_5^{(-)}$ can play the role of axion in the absence of $V_{\rm eff}[\gamma]$
in a similar type of five-dimensional model.\cite{Choi}
The equivalent model is reformulated in other formulation.\cite{G&W}
}
Hence we need to consider other candidates for the axion, the third solution or other (unkown) mechanism.
If we take the third solution, both $\mathcal{L}_{\rm brane}^{\theta}$ and $\mathcal{L}_{\rm bulk}^{\rm MCS}$ are forbidden
by a symmetry.
Based on the brane world scenario, let us relax the assumption such that 
the CP is an exact symmetry on the brane in an underlying high-energy theory, but it is not necessarily in the bulk.
Then we have the conditions:
\begin{eqnarray}
\frac{\kappa_{\tiny{\mbox{4D}}} \gamma}{g'_{\tiny{\mbox{4D}}}} < O(10^{-11})~,~~ {\mbox{argdet}}(M_u M_d) < O(10^{-10})~.
\label{C1}
\end{eqnarray}
The magnitude of $\mu$ parameter should be of order $O(1)$TeV from the naturalness argument
and it leads to the condition:
\begin{eqnarray}
q'_h \gamma = O(R/\mbox{TeV}^{-1})~.
\label{C2}
\end{eqnarray}
{}From the first condition of  (\ref{C1}) and (\ref{C2}),  the following relation is derived
\begin{eqnarray}
O(10^{14})\mbox{GeV} < \frac{g'_{\tiny{\mbox{4D}}}q'_h}{\kappa_{\tiny{\mbox{4D}}}} \frac{1}{R} ~.
\label{C3}
\end{eqnarray}

\section{Conclusions and discussion}

We have studied the generation of $\mu$ parameter from the dynamical rearrangement 
of local $U(1)$ symmetry in a five-dimensional model,
under the assumption that SUSY is broken by the Scherk-Schwarz mechanism.
We have obtained the same magnitude of $\mu$ parameter as that of soft SUSY breaking masses of hidden scalar fields (${\phi_k}^l$)
because the one-loop effective potential includes both the Wilson line phase and the Scherk-Schwarz phases of ${\phi_k}^l$.
We have discussed phenomenological implications on the $\theta$ parameter
and found that there can exist the constraint on the size of extra dimension in the absence of axion.

One of the problems in our scenario is that the origin of extra $U(1)$ symmetry is unknown.
There is an unusual feature that the eigenstates of $U(1)$ gauge symmetry do not agree with the eigenstates of BCs.
The origin of SM gauge singlets is also not clear.
It would be interesting to approach these problems from a more fundamental theory.

\section*{Acknowledgements}
This work was supported in part by scientific grants from the Ministry of Education, Culture,
Sports, Science and Technology under Grant Nos.~22540272 and 21244036 (Y.K.)
and No.~23$\cdot$9368 (T.M.).

\end{document}